\documentclass[aps,notitlepage,a4paper,superscriptaddress,11pt]{article}
\usepackage{multicol}
\usepackage[utf8]{inputenc}
\usepackage[dvips]{color}
\usepackage[toc]{appendix}
\usepackage[hang,flushmargin]{footmisc} 
\usepackage{titlesec,titletoc,ntheorem-hyper,authblk,abstract,latexsym,microtype,booktabs,amsmath,hyperref,cleveref,fancyhdr,wrapfig,array,amsfonts, amssymb,geometry,appendix,cite, secdot}
\usepackage{xcolor}

\newcommand{\be}{\begin{equation}}
\newcommand{\ee}{\end{equation}}
\newcommand{\bea}{\begin{eqnarray}}
\newcommand{\eea}{\end{eqnarray}}

\usepackage{graphicx,lipsum, cuted}
\usepackage{caption}
\usepackage{subfigure}
\usepackage{amsmath}
\usepackage{amsfonts}
\usepackage{amssymb}
\usepackage{stackengine}
\usepackage{mathtools}
\numberwithin{equation}{section}
\usepackage{titlesec}
\usepackage{sectsty}
\sectionfont{\bfseries\Large\raggedright}

\numberwithin{subcase}{case}
\usepackage{framed}
\allowdisplaybreaks
\usepackage[nottoc]{tocbibind}
\usepackage[free-standing-units]{siunitx}
\usepackage{helvet}
\usepackage{url}
\usepackage{titletoc}
\usepackage{blindtext}
\usepackage[gen]{eurosym}

\title{Tunneling of Hawking radiation for BTZ black hole revisited}

\author{Bijan Bagchi \\ 
e-mail: \href{mailto: bbagchi123@gmail.com}{bbagchi123@gmail.com}\\
\vspace{0.5cm}
Sauvik Sen \\
e-mail: \href{mailto: sauviksen.physics@gmail.com}{sauviksen.physics@gmail.com}} 
\affil{Department of Physics, School of Natural Sciences, Shiv Nadar University,\\ Gautam Buddha Nagar,
Uttar Pradesh 201314, India}
\date{November 2021}

\begin{document}

\maketitle

\begin{abstract}
    We re-examine Hawking radiation for a nonrotating (2+1)-dimensional BTZ black hole and evaluate the transmission probability of tunneling through the barrier of the event horizon employing the standard method of WKB approximation. Our results are presented for both uncharged and charged cases. We also explore the associated thermodynamics in terms of Hawking temperature and provide estimates of black hole parameters like the surface gravity and entropy.
\end{abstract}


\section{Introduction}

In the theory of general relativity (GTR) the black hole solutions are generally described by four types of metrics. The simplest one is of course the Schwarzschild metric with no electic charge and where there is spherical symmetry. To deal with the electric charge, Reissner–Nordstr\"{o}m metric is the relevant one that addresses the geometry of empty space surrounding a charged, nonrotating black hole. Kerr metric gives a generalization of the Schwarzschild metric wherein spin effects are taken into account. It was shown much earlier by Newman et al. \cite{new1, new2}
that the Kerr solution is derivable from the Schwarzschild solution for the vacuum Einstein case and further Kerr-Newman solution follows from the Reissner–Nordstr\"{o}m metric for the 
Einstein-Maxwell theory through a set of  complex-coordinate transformations in terms of a
null tetrad of basis vectors. In a wider perspective, Kerr-Newman metric deals with both charge and spin of the black hole (see, for a recent review, \cite{ada}).

To consolidate quantum gravity with other forces of nature one has to take into account extremely strong gravitational effects. A black hole is a natural place to look for them. Interest in (2+1)-dimensional gravity, which serves as a toy model for understanding how gravity operates in higher dimensions, owes to the foundational work of Ba\~{n}ados et al. (BTZ) who showed that such a simplified formulation had an embedded black hole solution \cite{ban}. 
In contrast 
to Schwarzschild and Kerr types of black holes, the BTZ is asymptotically anti-de Sitter (AdS) and not being asymptotically flat. The metric shows no curvature singularity at the origin. 
The static BTZ geometry is a solution of the Einstein equations with a fixed value of the negative cosmological constant $-1/l^2$ \cite{gaete, pantoja}. In \cite{ban} was included the cosmological constant possessing a negative
sign in the formalism of three-dimensional vacuum Einstein theory. 

For detailed analyses of general types of BTZ black holes having charge and spin see \cite{mar, clement2, akbar, sadeghi}.  Another interesting aspect of the BTZ black hole solution is that it reveals thermodynamical properties akin to the Schwardschild's black hole of (3+1)-dimensions. See \cite{carlip1} for an early review. BTZ black hole has been studied in the context of quantum gravity and Hawking temperature allowing for the presence of nonzero spacetime noncommutativity \cite{gup1}. For further related works see \cite{gup2, juric}.

In recent times, the construction of AdS charged BTZ black hole was carried out and a class of plausible solutions were explored in \cite{hen}. The dynamical and thermodynamical behaviour were investigated and Hawking radiation \cite{haw} was estimated by Parikh and Wilczek \cite{par} in the background  of conventional quantum mechanical tunneling (see also \cite{kim2, yale, noz}). It needs to be emphasized that the study of Hawking radiation in the context of BTZ black hole and null-geodesic formalism within the scenario of \cite{par} was also taken up by Wu and Jiang \cite{swu} for the rotational uncharged BTZ black hole. An idea to measure Hawking-like features and the Unruh effect in the laboratory was proposed in \cite{ros}. Furthermore, the effect of generalized uncertainty principle was inquired to assess Hawking radiation in (2+1)-dimensions \cite{babar}. In addition, the effect of quantum gravity on the stability aspects of black holes was discussed in \cite{ali}. The BTZ black hole was also analyzed for massive gravity in a more general context \cite{dey} to ascertain Hawking radiation and provide an estimate of tunneling probability. The particular work of Kim \cite{kim} is especially interesting in that he studied viable black hole solutions in the context of its global structure \cite{hawk}.  In a similar spirit he attempted a derivation as in \cite{new1, new2} but concentrated only for the three-dimensional spacetime. 
His analysis rested on the interpretation that the three-dimensional geometry is just a $\theta = \frac{\pi}{2}$ cut of the four-dimensional spacetime with $\theta$ representing the polar angle.
In particular, his nonrotating result correctly resembled the BTZ solution.

One of the purposes of this work is to use the BTZ metric to reassess the transmission probability of radiation for nonrotating black holes in (2+1)-dimensions corresponding to the uncharged and charged cases. We remark that apart from [17], study of tunneling was also taken up for the dilatonic black hole in \cite{hu}. However, as will be clear from the steps provided in the following section, the approach and analysis of the present work is quite different. Noting that the picture for the (2+1)-dimension is markedly different from the (3+1)-case in that as per our study the temperature is found to be proportional to the square root of mass rather than inverse of it for both uncharged and charged black holes \cite{carlip1}, we demonstrate that the tunneling probability for the charged black holes emerges smaller than the uncharged one. Interestingly, our finding is in tune with what is obtained in  \cite{gup1} by including noncommutative effects when rotational effect is ignored.

In Section II, we set up the metric by transforming the Painlev\'{e} time to Schwarzschild time with the help of a suitable shift involving an arbitrary function of the radial coordinate. This facilitates arriving at the generalized structure of the metric for the (2+1)-dimensional black hole. It is then exploited to address the issue of the tunneling probability of Hawking radiation for an uncharged black hole employing the WKB approximation and comparing the result with the existing estimates made from different perspectives. In section III, we consider in a similar way the case of the charged black hole. In Section IV, we compute the essential thermodynamic quantities \cite{pra} like the surface gravity and entropy of a black hole. Finally, in section V, we present a short summary.

\section{Tunneling for uncharged nonrotating BTZ black hole}

We begin by representing the BTZ metric for a nonrotating uncharged black hole of mass M in spherical polar coordinates by 

\begin{equation}
    ds^2 = Z(r) dt^2 - Z^{-1}(r) dr^2 - r^2 d\phi^2
\end{equation}

\begin{equation}
    Z(r) = -M + \displaystyle\frac{r^2}{l^2}
\end{equation}
Throughout this work we will employ natural units. In (2.1) $t$ is the Painlevé time and we have followed Kim's \cite{kim} treatment by picking the slice  $\theta = \pi/2$. As noted by Pantoja et al \cite{pan}, the metric (2.1) is a solution of the vacuum Einstein field equations taking the cosmological
constant value to be $\Lambda = - \frac{1}{l^2}$. As elaborated there, corresponding to a strictly positive mass value, the horizon of the black hole is at the point $r = \sqrt{M} l$, for a negative mass value, the black hole is non-existent while the interval $-1 < M < 0$ is linked to the metric generated by a point source at the origin. The vacuum state corresponds to $M = 0$. In the present problem, because of $M>0$, the positive values of the outer $(r_{out})$ and inner horizon $(r_{in})$ are automatically obeyed.

Without loss of generality, let us transform to the Schwarzschild time ($t_{Sch}$) by making the shift 

\begin{equation}
    t =t_{Sch} - \xi(r)
\end{equation}
where $\xi(r)$ is an arbitrary function of r and the prime denotes the derivative with respect to r. Accordingly, $dt^2$ becomes

\begin{equation}
        dt^2 = dt_{Sch}^2 + \xi'^2(r)dr^2-2\xi'(r)drdt_{Sch}
\end{equation}

Substituting (2.4) in (2.1) we derive the following form for $ds^2$
\begin{equation}
\begin{split}
ds^2 = Z(r) dt^2_{Sch} + [Z(r) \xi'^{2}(r) - Z^{-1}(r)] dr^2 
- 2 Z(r)\xi'(r)dr dt_{Sch} - r^2d\phi^2
\end{split}
\end{equation}
Since $\xi(r)$ is arbitrary we choose it to render the coefficient of $dr^2$ unity. Thus we are led to the following pair of relations
\begin{equation}
    \begin{split}
        Z(r)\xi'(r)^2-Z^{-1}(r)=1 \hspace{2mm}\longrightarrow \hspace{2mm} \xi'(r) = \pm\sqrt{Z^{-1}(r)+Z^{-2}(r)}
    \end{split}
\end{equation}
We can therefore recast (2.5) in the form

\begin{equation}
\begin{split}
   ds^2 = Z(r) dt^2_{Sch} + dr^2 - r^2d\phi^2  -2Z(r)\sqrt{Z^{-1}(r)+Z^{-2}(r)}dr dt_{Sch} 
   \end{split}
   \end{equation}
The radial null geodesic will obey

\begin{equation}
\begin{split}
 Z(r)dt_{Sch}^2+dr^2-2Z(r)
 \sqrt{Z^{-1}(r)+Z^{-2}(r)}drdt_{Sch} =0
 \end{split}
\end{equation}
and hence we have

\begin{equation}
\begin{split}
 dt_{Sch}^2\Big(Z(r)+\big(\displaystyle\frac{dr}{dt_{Sch}}\big)^2 -2Z(r)\sqrt{Z^{-1}(r)+Z^{-2}(r)}
 \big(\displaystyle\frac{dr}{dt_{Sch}}\big)\Big) =0
 \end{split}
\end{equation}
(2.9) can be rewritten in the form

\begin{eqnarray}
&& \dot{r}^2-2\dot{r}Z(r)\sqrt{Z^{-1}(r)+Z^{-2}(r)}+Z(r) =0 \hspace{2mm}\longrightarrow \hspace{2mm} \dot{r}=\sqrt{1+Z(r)}\pm 1
\end{eqnarray}
where the overdot refers to the derivative with respect to the Schwarzschild time. The negative sign refers to an incoming geodesic while the positive sign stands for an outgoing geodesic. It is assumed that the pair production occurs just inside the event horizon at $r_{in}\approx r_+$.

A couple of points are in order. If $\Omega$ is the energy of the particle that is created during the pair production then the mass of the black hole after emission would be $M-\Omega (>0)$. Since $r_{in}$ is the radius of the event horizon when the particle is produced just inside it, while $r_{out}$ is the radius of the event horizon when the particle has tunneled across and emitted out of the black hole, these are determined from solutions of $Z(r) = 0$
\begin{align}
\begin{split}
        r_{in}&=l\sqrt{M}\\
        r_{out}&=l\sqrt{M-\Omega}
\end{split}
\end{align}
where $r_{out}<r_{in}$, with $r_{in}$ in conformity with the constraint for a positive mass mentioned earlier. With this background we proceed to calculate the tunneling probability of Hawking radiation.

The region between $r_{in}$ and $r_{out}$ separating the two points acts as a potential barrier for the tunneling particle to overcome. For the particle produced during pair production, its energy would be lower than the acting barrier thus mimicking a classically forbidden region. In this region the action $\zeta$ is imaginary and we can profitably use the WKB approximation, as a semiclassical way, to estimate the tunneling probability \cite{par, dey}. See the works in  \cite{feng1, feng2, feng3, fla, li, fle} where the issue of tunneling probability has also been addressed.
Indeed writing for the imaginary part of the action
\begin{equation}
    \mbox{Im}\zeta=\mbox{Im}\int_{r_{in}}^{r_{out}}p_rdr = \mbox{Im}\int_{r_{in}}^{r_{out}}\int_{0}^{p_r}dp_rdr
\end{equation}
and noting from Hamilton's equation the relationship
\begin{equation}
     dp_r=\displaystyle\frac{dH}{\dot{r}}
\end{equation}
where $H$ takes the values $M$ and $M - \Omega$ corresponding to $p_r=0$  and $p_r=p_r$ respectively, we project the integral (2.12) to read
\begin{equation}
    \mbox{Im}\zeta= \mbox{Im}\int_{r_{in}}^{r_{out}}\int_{M}^{M-\Omega}\displaystyle\frac{dH}{\dot{r}}dr
\end{equation}
Then, substituting the value of $\dot{r}$ from  {(2.10)} and concentrating on the outgoing geodesic we are led to 

\begin{equation}
    \mbox{Im}\zeta= \mbox{Im}\int_{r_{in}}^{r_{out}}\int_{M}^{M-\Omega}\displaystyle\frac{dH}{1+\sqrt{1+Z(r)}}dr
\end{equation}
Let $\mathcal{M}_{in}(=M)$ and $\mathcal{M}_{out}(=M-\Omega)$ be the respective mass of the black hole before and after the pair production. Since the emitted particle has a very negligible mass, we can approximate 
\begin{align}
    \begin{split}
        \mathcal{M} &= \mathcal{M}_{\mbox{in}} \approx \mathcal{M}_{\mbox{out}}\\
        d \mathcal{M} & \approx -d\Omega
    \end{split}
\end{align}
The integral limits in (2.15) translate to $\Omega=0$ when $\mathcal{M}=M$ and $\Omega=\Omega$ when $\mathcal{M}=M-\Omega$ with $dH$ replaced by $-d\Omega$. As such, the expression (2.15) becomes 
\begin{equation}
    \mbox{Im}\zeta= -\mbox{Im}\int_{r_{in}}^{r_{out}}\int_{0}^{\Omega}\displaystyle\frac{d\Omega}{1+\sqrt{1+Z^(r)}]}dr
\end{equation}
From the consideration that the emitted particle has a very small energy of the order of $\Omega$, binomial expansion in the vicinity of the event horizon, where the quantity $Z$ is supposed to be small, gives a simplified representation
\begin{equation}
     \mbox{Im}\zeta= -\mbox{Im}\int_{r_{in}}^{r_{out}}\int_{0}^{\Omega}\displaystyle\frac{2d\Omega}{4+Z(r)}dr
\end{equation}
to first order in $Z$. Replacing $M$ by $M-\Omega$ and substituting (2.2) in (2.18) we obtain
\begin{equation}
    \mbox{Im}\zeta= -\mbox{Im}\int_{r_{in}}^{r_{out}}\int_{0}^{\Omega}\displaystyle\frac{2d\Omega}{4+ \left [-M+\Omega+\displaystyle\frac{r^2}{l^2} \right ]}dr
\end{equation}
To tackle the above integral we substitute $-M+\Omega+\displaystyle\frac{r^2}{l^2} = -4\alpha$  implying $d\Omega = -4 d\alpha$. Further, the limits
$\Omega=0 \rightarrow \alpha=\alpha(0)$ and $\Omega= \Omega \rightarrow \alpha=\alpha(\Omega)$ also hold. As a consequence, $\mbox{Im}\zeta$ is convertible to the form

\begin{equation}
     \mbox{Im}\zeta= 2\hspace{1mm}\mbox{Im}\int_{r_{in}}^{r_{out}}\int_{\alpha(0)}^{\alpha(\Omega)}\displaystyle\frac{d\alpha}{1-\alpha}dr
\end{equation}
We now use the residue theorem of complex analysis to estimate the integral (2.20). Noticing that there is a simple pole at $\alpha=1$ and that the residue at this point is $-1$ it follows that 
\begin{equation}
\mbox{Im}\zeta =-4\pi(r_{out}-r_{in})=4\pi(r_{in}-r_{out})
\end{equation}
This gives straightforwardly the following result for the tunneling probability
\begin{equation}
\Gamma(\Omega) \approx \exp(-2\mbox{Im}\zeta) = \exp(-8\pi\rho)
\end{equation}
where $\rho = r_{in}-r_{out} = \displaystyle\frac{\Omega l}{2\sqrt{M}} + \mathcal{O}(\Omega^2)$. A comparison of the right side of (2.21) with the estimate made in \cite{swu} for the rotational uncharged BTZ black hole shows that ours is a factor of 2 too large. More precisely, while (2.22) shows the dependence of $\exp(-\frac{1}{\sqrt{M}})$, the $(3+1)$-dimensional calculation of Parikh and Wilczek \cite{par} depicts a slightly different behaviour of $\exp(-M)$. A similar form like (2.21) was obtained in the massive gravity case \cite {dey} but the difference of $r_{in}$ and $r_{out}$ depended on a complicated function of the mass. It is interesting to note that if rotational contributions are discarded then the noncommutative result of \cite{gup1} matches with ours including the dependence of $\Gamma$ on $\exp(-\frac{1}{\sqrt{M}})$.

\section{Tunneling for charged nonrotating BTZ black hole}

The charged counterpart of the nonrotating BTZ metric \cite{sadeghi, mar, pra, akbar } can be written as:

\begin{equation}
    ds^2 = Y(r) dt^2 - Y^{-1}(r) dr^2 - r^2 d\phi^2
\end{equation}

\begin{equation}
    Y(r) = -M + \displaystyle\frac{r^2}{l^2} - \displaystyle\frac{\pi}{2}Q^2 \mbox{ln}r
\end{equation}

We follow a similar prescription as in the previous section to calculate the tunneling probability of this black hole.
The radius of the event horizon before and after emission are given by .

\begin{equation}
\begin{split}
     r_{in} = \displaystyle\frac{\pi Q^2 l^2}{4}+\displaystyle\frac{1}{4}\sqrt{ \pi^2 Q^4 l^4 - 8 \pi Q^2 l^2 + 16 l^2 M}\qquad \quad \\
       r_{out} = \displaystyle\frac{\pi Q^2 l^2}{4}+\displaystyle\frac{1}{4}\sqrt{\pi^2 Q^4 l^4 - 8 \pi Q^2 l^2 + 16 l^2 (M-\Omega)}
       \end{split}
       \end{equation}
This is obtained by setting $Y(r)=0$ similar to the vanishing criterion of $Z(r)$ undertaken in the previous section to determine the corresponding $r_{in}$ and $r_{out}$.\\
The imaginary action can then be expressed as:

\begin{equation}
\mbox{Im}\zeta = -\mbox{Im}\int_{r_{in}}^{r_{out}}\int_0^\Omega
\displaystyle\frac{2d \Omega}{4+[-(M-\Omega)+\displaystyle\frac{r^2}{l^2}-\displaystyle\frac{\pi}{2}Q^2\mbox{ln}r]}dr
\end{equation}

After suitable substitution and using the residue theorem to solve the above integral, the final tunneling probability for the charged BTZ black hole is determined to be
\begin{equation}
    \Gamma(\Omega) \approx \exp(-8\pi\sigma) 
\end{equation}
where 
\begin{equation}
\sigma = r_{in}-r_{out} = \displaystyle\frac{2l^2\Omega}{\sqrt{\pi^2 Q^4 l^4 - 8 \pi Q^2 l^2 + 16 l^2 M}} + \mathcal{O}(\Omega^2) 
\end{equation}
and $\pi^2 Q^4 l^4 + 16 Ml^2 > 8\pi Q^2 l^2$. Comparing with the result (2.22) we see that the tunneling probability for the charged nonrotating black hole is smaller than the uncharged one. Further, we remark that the dependence on the mass M of the argument of the above exponential is different from the $(3+1)$-dimensional case \cite{par}, where it was found that for a black hole with charge Q the transmission probability went like $\approx \exp \left [-4\pi (2\Omega(M-\frac{\Omega}{2}))-(M-\Omega)\sqrt{(M-\Omega)^2-Q^2}+M\sqrt{M^2-Q^2} \right ]$.

\section{Thermodynamical aspects} 

The influential works of Bekenstein, Hawking, Bardeen et al \cite{beken, haw, bardeen} in the 1970s paved the way for active research in black hole thermodynamics. In particular, Bekenstein \cite{beken} postulated that the area of the event horizon was proportional to its entropy. Further works by Bardeen et al \cite{bardeen} laid the foundations of the four laws of black hole thermodynamics by drawing an analogy from the conventional ones of thermodynamics. 

In the present context we set up our results against the standard ones of \cite{carlip2, kubiznak, frassino}.
According to Hawking's theory \cite{haw} the so-called information paradox could be resolved by the emission of particles from a black hole due to quantum pair-production near the event horizon. One denotes by $T_H$ as characteristic temperature when the radiation occurs. The value of $\rho$ obtained for the uncharged black hole when substituted in (2.22) gives the Boltzmann factor $\exp(-{\frac{\mathcal{E}}{T_H}})$ where $\mathcal{E} = \Omega$ is the energy of the emitted particle. Similarly for the charged case we employ the estimate of $\sigma$ made in (3.6). This provides us with the following expression of Hawking temperature ($\mathcal{T}$) of the (2+1)-dimensional uncharged (uc) and charged (c) black hole:

\begin{equation}
    \mathcal{T}^{uc} = \displaystyle\frac{\sqrt{M}}{4\pi l}
\end{equation}
\begin{equation}
    \mathcal{T}^c = \displaystyle\frac{\sqrt{\pi^2 Q^4 l^4 - 8 \pi Q^2 l^2 + 16 l^2 M}}{16\pi l^2}
\end{equation}
Since the surface gravity $\kappa$ in both cases is $2\pi \mathcal{T}$ we at once deduce straightforwardly
\begin{eqnarray}
 && \kappa^{uc} = \frac{\sqrt{M}}{2l}\\
 && \kappa^{c} =  \displaystyle\frac{\sqrt{\pi^2 Q^4 l^4 - 8 \pi Q^2 l^2 + 16 l^2 M}}{8l^2}
\end{eqnarray}

We know from Bekenstein's work \cite{beken} that the area of a black hole event horizon is proportional to its entropy. This means that just like entropy can never decrease, the area of the event horizon of a black hole can only increase. In (3+1)-dimensions the surface of a black hole is the area of a sphere ($4\pi r^2$). In $(2+1)$ dimensions the analogue can be identified with its perimeter $A=2 \pi r_{out}$. A simple derivation yields for the respective areas of uncharged and charged black holes
\begin{eqnarray}
  &&  A^{uc} = 2\pi l \sqrt{M-\Omega} \\
&& A^{c} = \displaystyle\frac{\pi^2 Q^2 l^2}{2}+ \displaystyle\frac{\pi}{2}\sqrt{\pi^2 Q^4 l^4 - 8 \pi Q^2 l^2 + 16 l^2 (M-\Omega)} 
\end{eqnarray}
These relations imply for the entropy relations

\begin{eqnarray}
 &&   S^{uc} = \displaystyle\frac{A}{4} =\displaystyle\frac{\pi l}{2}\sqrt{M-\Omega} \\
 && S^{c} = \displaystyle\frac{A}{4} =\displaystyle\frac{\pi^2 Q^2 l^2}{8}+\displaystyle\frac{\pi}{8}\sqrt{\pi^2 Q^4 l^4 - 8 \pi Q^2 l^2 + 16 l^2 (M-\Omega)} 
\end{eqnarray}
Note that the functional forms of $S^c$ and $S^{uc}$ are different. Evidently, because of the influence of the charge $Q$, $S^c$ is higher than $S^{uc}$.

\section{Summary}

In this paper we introduced a set of complex coordinate transformations to reinterpret the BTZ metric in terms of Painlev\'{e} coordinates. Thus the basic equations were set up that give conditions for the incoming and outgoing geodesics. Following Parikh and Wilczek's approach, we then evaluated the transmission probability of Hawking radiation for both uncharged and charged nonrotating black holes in the framework of WKB approximation. The related issue of black hole thermodynamics was also discussed and estimates of typical thermodynamical parameters, like the surface gravity and entropy, were made.

\section{Acknowledgment}

One of us (SS) thanks Ricardo Troncoso and Tajron Juri\'c for enlightening correspondences. He also acknowledges support from Shiv Nadar University for a research fellowship. Both of us are indebted to the anonymous reviewer for carefully going through our manuscript and advising a number of constructive changes.


\begin{thebibliography}{26}

\bibitem{new1}  E. T. Newman and A. I. Janis, J. Math. Phys. \textbf{6}, 915 (1965). 

\bibitem{new2}  E.
T. Newman, E. Couch, R. Chinnapared, A. Exton, A. Prakash,
and R. Torrence, J. Math. Phys. \textbf{6}, 918 (1965).

\bibitem{ada}  T. Adamo and E. T. Newman, Scholarpedia \textbf{9(10)}, 31791 (2014). 

\bibitem{ban} M. Ba\~{n}ados, C. Teitelboim, and J. Zanelli, Phys. Rev. Lett. \textbf{69},  1849 (1992).

\bibitem{gaete} M. Bravo-Gaete and M. Hassaine, Phys. Rev. \textbf{D 90}, 024008 (2014).

\bibitem{pantoja} N. R. Pantoja, H. Rago and R. O. Rodriguez, J. Math. Phys. \textbf{45}, 1994 (2004).


\bibitem{clement2} G. Cl\'ement, Phys. Lett. B. \textbf{367}, 70 (1996).

\bibitem{sadeghi} J. Sadeghi and V. R. Shajiee, Eur. Phys. J. Plus \textbf{132}, 1-5 (2017).



\bibitem{mar} C. Martinez, C. Teitelboim, and J. Zanelli, Phys. Rev. \textbf{D 61}, 104013 (2000).

\bibitem{akbar} M. Akbar et al, Phys. Rev. \textbf{D 83}, 084031 (2011).



\bibitem{carlip1} S. Carlip, Class. Quant. Grav. \textbf{12}, 2853 (1995).



\bibitem{gup1} K. S. Gupta, T. Juri\'c and A. Samsarov, J. High Eng. Phys. \textbf{2017}, 107 (2017).

\bibitem{gup2} K. S. Gupta, E. Harikumar, T. Juri\'c, S. Meljanac and A. Samsarov, Adv. High Eng. Phys. \textbf{2014}, 139172 (2014).

\bibitem{juric} T. Juri\'c and A. Samsarov, Phys. Rev. \textbf{D 93}, 104033 (2016). 

\bibitem{hen} S. H. Hendi, B.E. Panah, and S. Panahiyan, J. High Energy Phys. \textbf{2016}, 029 (2016).

\bibitem{haw} S. W. Hawking, Commun. Math. Phys. \textbf{43}, 220 (1975); Nature \textbf{248}, 5443 (1974).

\bibitem{par} M. K. Parikh and F. Wilczek, Phys. Rev. Lett. \textbf{85}, 5042 (2000).

\bibitem{kim2} H. Kim, Phys. Lett. \textbf{B 703}, 94 (2011).

\bibitem{yale} A. Yale, Phys. Lett. \textbf{B 697}, 398 (2011).

\bibitem{noz} K. Nozari and S. H. Mehdipour, Class. Quantum Grav. \textbf{25}, 175015 (2008).

\bibitem{swu} S-Q Wu and Q-Q Jiang, J. High Eng. Phys. \textbf{03}, 079 (2006).



\bibitem{ros} H. C. Rosu, Int. J. Mod. Phys. \textbf{D03}, 543 (1994).

\bibitem{babar} R. Babar, W. Javed and A. \"{O}vg\"{u}n, Mod. Phys. Lett. \textbf{35}, 2050104 (2020).

\bibitem{ali} R. Ali, K. Bamba and S. A. A. Shah, Symmetry \textbf{11}, 631 (2019).

\bibitem{dey} S. Chougule,
S. Dey, B. Pourhassan, and M. Faizal, Eur. Phys. J. \textbf{C 78}, 685 (2018).

\bibitem{kim} H. Kim, Phys. Rev. \textbf{D 59}, 064002 (1999).


\bibitem{hawk} S. W. Hawking, C. J. Hunter, and M. M. Taylor-Robinson, Phys. Rev. \textbf{D 59}, 064005 (1999).

\bibitem{hu} Y. Hu, J. Zhang and Z. Zhao, Int. J. Mod. Phys. \textbf{16}, 847 (2007).

\bibitem{pra} P. Prasia and V. C. Kuriakose, Eur. Phys. J. \textbf{77C}, 27 (2017).

\bibitem{pan} N. R. Pantoja, H. Rago and R. O. Rodr\'{i}guez,  J. Math. Phys. \textbf{45}, 1994 (2004).

\bibitem{feng1} Z. W. Feng, H. L. Li, X. T. Zu, S. Z. Yang, European Physical Journal C76, 212 (2016).

\bibitem{feng2} Z. W. Feng, Q. C. Ding, S. Z. Yang, European Physical Journal C 79, 445 (2019).

\bibitem{feng3}  Z. W. Feng, X. Zhou, S. Q. Zhou, D. D. Feng, Annals of Physics, 416, 168144 (2020).

\bibitem{fla}  E. E Flanagan, Phys. Rev. Lett. 127, 041301.

\bibitem{li}  R. Li, J. Wang, Phys. Rev. D 104, 026011 (2021).

\bibitem{fle} C. H. Fleming, Hawking radiation as tunneling, \url{http://www.physics.umd.edu/grt/taj/776b/fleming.pdf}. 



\bibitem{beken} J. D. Bekenstein Nuovo Cim. Lett. \textbf{4}, 737-740 (1972);
Phys. Rev. \textbf{D 7}, 2333-2346 (1973); Phys. Rev. \textbf{D 9}, 3292 (1974).


\bibitem{bardeen} J. M. Bardeen, B. Carter, and S. W. Hawking, Commun. Math. Phys. \textbf{31}, 161-170 (1973).

\bibitem{carlip2} S. Carlip, Phys. Rev. \textbf{D 51}, 632 (1995).

\bibitem{kubiznak} D. Kubizňák, R. B. Mann, and M. Teo, Class. Quantum Grav. \textbf{34}, 063001 (2017).

\bibitem{frassino} A. M. Frassino, R. B. Mann, and J. R. Mureika, Phys. Rev. \textbf{D 92}, 124069 (2015).





 \end{thebibliography}
 \end{document}